\documentstyle{article}
\include{epsf}
\begin{document}

\title{ Generalized (m,k)-Zipf law for fractional Brownian motion-like time
series with or without effect of an additional linear trend }

\author{ Ph. Bronlet and M. Ausloos\\
SUPRAS and GRASP, B5, Sart Tilman Campus, \\B-4000 Li$\grave e$ge,
Euroland}

\date{\today}

\maketitle

\begin{abstract} { We have translated fractional Brownian motion (FBM) signals
into a text based on two ''letters'',  as if the signal fluctuations correspond
to a constant stepsize random walk.  We have applied the Zipf method to extract
the $\zeta '$ exponent relating the word frequency and its rank on a log-log
plot. We have studied the variation of the Zipf exponent(s) giving the
relationship between the frequency of occurrence of words of length
$m<8$ made of
such two letters: $\zeta '$ is varying as a power law in terms of $m$. We have
also searched how the $\zeta '$ exponent of the Zipf law is influenced by a
linear trend and the resulting effect of its slope. We can distinguish finite
size effects, and results depending whether the starting FBM is persistent or
not, i.e. depending on the FBM Hurst exponent $H$.  It seems then numerically
proven that  the Zipf exponent of a persistent signal  is more
influenced by the
trend than that of an antipersistent signal. It appears that the
conjectured law
$\zeta ' = |2H-1|$ only holds near $H=0.5$. We have also introduced
considerations based on the notion of a {\it time dependent Zipf law} along the
signal. }

\end{abstract} \vskip 0.5cm


\vskip 0.5cm

\section{Introduction}

Many phenomena which contain discernible events which can be counted can be
ranked according to their frequency, and a so called Zipf plot can be drawn.
\cite{Zipf,West,Hill1,Hill2,Montemurro} Very often a  $quasi$ linear
relationship
is found on  a log-log plot. The slope corresponds to an exponent $s$
describing
the frequency $P$ of the $cumulative$ occurrence of the events
according to their
rank $R$ through, e.g. $P(>R) \simeq R^{-s}$.

Such a ($Zipf$) $power$ $law$ is found in many cases, with $s \sim 1$: see the
distribution of $income$ of individuals or companies in countries (Pareto
distribution) \cite{Pareto,MandelPareto,Okuyama}, in economy with the $size$ of
companies \cite{companysize,Ramsden},  in $earthquakes$ (Gutenberg-Richter law)
\cite{GRlaw,SKKV}, in $city$ distribution \cite{MarsiliZhang,Gabaix}, etc.
\cite{Ma,Piqueira,Li,Kalda,z1,z2,Ausloosexotic,domino}...  The Zipf
law universal
feature is thought to originate from stochastic processes \cite{CSV}, in
particular when they can be modeled as random walks in a log scale
\cite{Kawamura}, - though it is still often said in a lay language that the law
is a description ({\it or a result ?}) of uniformity and diversity.

A simple extension of the Zipf analysis is to consider $m$-letter
words, i.e. the
words strictly made of $m$ characters without considering the white spaces. The
available number of different letters or characters $k$ in the alphabet should
also be specified. A power law for the word frequency $f(R)$ is expected to be
observed \cite{z1,z2}

\begin{equation} f~ \sim ~R^{-\zeta}.  \end{equation}

The Zipf exponent can be estimated through the derivative of the best
linear fit
on a log-log graph. There is no theory at this time predicting whether the
exponent is a function of ($m,k$). Elsewhere\cite{z2} we had already shown that
the Zipf exponent could be different, call it $\zeta '$  if  the
frequency $f$ of
occurrence is normalized with respect to the theoretical frequency $f'$ which
should be that expected for pure, or unbiased, (stochastic) Brownian processes,
thus

\begin{equation}  f/f'  ~ \sim~ R^{-\zeta '} . \end{equation}

E.g.  suppose a binary alphabet, i.e.  made up of two characters,
$u,d$ (or 0 and
1 in electronics)  with occurrence $p_u$ and $p_d$; the theoretical frequency
$f'$ for the  number $n$ of characters, say of the type $d$ in a word of length
$m$ is $f'=p_{u}^{(m-n)}p_{d}^{(n)}$.

Previous Zipf-like analyses have often neglected the possibility that
the ($u,d$)
distribution could be biased, over a finite size time interval, thus a finite
tendency might exist, - and in fact could have a quite varied
structure, as  when
in finance a trend is considered to be mimicked by some moving average
\cite{movaverage}. It seems clear that if a tendency is positive the number of
positive volatility values is larger than the number of negative ones (and
conversely) \cite{Ausloosexotic}. Therefore the Zipf law exponent might be
affected because of some bias in the ranking. Whether or not the
$\zeta$ exponent
depends on the bias is briefly examined here.

The final goal of this series of investigations is to apply the idea
in the study
of financial temporal series, or generally translating a time series into a
sentence, a particular letter corresponding to a particular variation of a
signal. This is in line with previous studies in econophysics e.g. in order to
search whether some investment strategy can be derived in particular
from a Zipf
law observation. This would fall into the same type of studies as
those implying
the detrended fluctuation analysis (DFA). \cite{Ausloosexotic}

Here we analyze time series based on a one dimensional fractional
Brownian motion
\cite{MandelPareto,Addison} characterized by the so called Hurst exponent $H$.
\cite{Addison,Hurst}

Why such a series? Because the FBM is not a Markovian process since
its value at
a given time depends on all past points, whence we consider that it can be a
useful model for modeling financial data time series.  In fact, Peters
\cite{Peters1,Peters2} has shown that a FBM is a good model for describing {\it
returns} (but it does not work for $options$) in financial series.

\section{Data}

In order to develop a FBM we have used  Rambaldia and Pinazza algorithm
\cite{RP}. According to the latter a FBM can be obtained from \begin{equation}
B_H(j)=\sum_{i=1}^j \omega_{j-i+1}\xi_i , \end{equation} where $\xi$ represents
the "walker" position during a time interval $\Delta t$; $\xi$ is a random
variable to be extracted from a Gaussian distribution with zero mean
and a given
variance. Thereby the signal is a stochastic one, with a diffusion growing with
time as $2H$.\cite{stochdiff} The weight function $\omega_{j-i+1}$ is given by

\begin{equation}
\omega_{j-i+1}=\frac{\gamma}{H+\frac{1}{2}}[(j-i+1)^{H+\frac{1}{2}}-(j
-i)^{H+\frac{1}{2}}]
, \end{equation} where $\gamma$  is such that $<B_H^2(1)>=1$. If
$H=1/2$ one has
the usual Brownian motion. The signal is said to be persistent for $H>1/2$, and
antipersistent otherwise.  There has been some conjecture \cite{czirok,troll} ,
sometimes thought to be proven like a theorem that \begin{equation} \zeta ' =
\zeta = |2H-1|. \end{equation}

We have created six different FBM signals with $H$ values
respectively equal to
0.17, 0.41, 0.47, 0.60, 0.67, 0.82. The series have a 16 384 ( = 2$^{14}$)
length.  They are normalized such that

\begin{equation} B_H(t)= \frac{B_{H}(t)-1.1B_{H,min}(t)}{1.1 B_{H,Max}(t)}.
\end{equation}

The coefficient 1.1 in the denominator  and numerator is used in order to avoid
zero and unity values. The series are  shown in Fig.1. Their characteristic is
summarized in Table 1. We have recalculated the Hurst exponent \cite{Hurst} by
the box counting method.\cite{Addison} The error bar is given in Table 1, as
$\Delta H$. The error bars are those resulting from a root mean
square analysis.
The linear trend has been measured and is reported to be of the order of
$10^{-5}$, obviously due to the finite size of the system.

In order to apply the Zipf method and extract the $\zeta '$ exponent
relating the
word frequency and its rank, we have translated the FBM signals into
a text based
on two $letters$, $u$ and $d$, occurring with a frequency  $p_u$ and $p_d$
respectively.   The bias defined as  $\epsilon = p_u -0.5$ has been
measured. The
bias and also the signal tendency have been observed as a function of time but
are not shown here for lack of space. They decay  quickly, can be positive or
negative and are  of the order or smaller than $1 \%$ after 4000 time steps.

The partial distribution functions (pdf) of the logarithm of the signal
volatility , i.e. \begin{equation} Z(t)=\ln(y(t+\Delta t))-\ln(y(t))
\end{equation} have been fitted to Gauss and stretched exponential
distributions.
Instead of ranking the $Z(t)$ values in constant size box histograms,
one can as
suggested by Adamic \cite{Adamic} use binned histograms with bin size
increasing
exponentially, thereby obtaining other {\it  best fit parameters}.   The
stretched exponential distribution seems to well describe the signal
fluctuations. A Kolomogorov-Smirnov test has also been made. \cite{KS}   Notice
that even though the KS distance increases when the exponential size
box is used,
this scheme reduces the uncertainty values and is found to lead to better fits.

Another test of the stochasticity (or not) of the data is based on
the surrogate
data method \cite{surrogate} in which one randomizes either the sign of the
fluctuations or shuffles their amplitude and Fourier transform the resulting
signal  in order to observe whether a white noise signal is so
obtained. Finally
we have observed whether the error bars (or confidence intervals) of the raw
signal and the surrogate data signal (not shown here) overlap.  The
characteristic of the spectral functions $S(f) \sim f ^{-\beta}$ so
obtained are
available from the authors if necessary.  The results allow us to conclude that
the above FBM signals are satisfactory for further treatment in
presence of a to
be pre-imposed bias.

Notice that similar histograms of such "words" were already
published\cite{Zhang,MolgedeyEbeling} for  ($m=3$, $k=2$) and ($ m=5$, $k=3$)
respectively, but the authors were more interested in deviations from
randomness
than in the Zipf exponent.

\section{Zipf Analysis of Fractional Brownian Motion Raw Data}

Consider the ($m$,2) Zipf method, thus for an only two character alphabet, and
words of arbitrary length $m$. Therefore there are $2^m$ different possible
words.  We searched whether these words exist in the series of Fig.1, counted
them and ranked them in decreasing order, as shown in Fig. 2, for $2 \le m \le
8$. The $\zeta '$ (and $\zeta $, also but not shown) exponents seem to increase
with $m$ (Fig. 3). The result may be attributed to  the finite size of the
series, if one realizes that for such series the number of long words
necessarily
occurs  more rarely than the number of short ones.

As a test of finite size effects, we have successively removed one by one the
less frequently occurring words, and recalculated the $\zeta '$ (and $\zeta $)
exponents, in some sense taking a {\it rank}  $\rightarrow$ zero
limit, or first
derivative. The exponents resulting from the average of the latter values are
shown in Fig. 4. It is found that for large $H$, i.e. a persistent signal,  the
exponents are rather constant, but still markedly vary for the antipersistent
signals.

As a  further step, we can also consider whether there is a Zipf law
$evolution$,
i.e. search for the $\zeta '$ exponent evolutions, as a function of time $t$ or
as a function of the number of points in the series.  We have calculated values
of $\zeta '$ and $\zeta $ for the (first) box containing 100 points, then for a
box containing 200, 300, ...  etc  points up to 16 000. The evolution
(Fig. 5) is
rather drastic for the first, say, 6000 points but is moderately stable
thereafter.

Next recall the so called local (or better $instantaneous$) DFA method.
\cite{Ausloosexotic,Azbel,DNA,v1,a2,m1,m2} In order to probe the existence of
{\it correlated and/or decorrelated sequences}, a so-called observation box  of
finite size can be constructed, and is placed at the beginning of the
data. A DFA
is performed on the data contained in that box. The box is then moved along the
historical time axis by a few points toward the right along the sequence.
Iterating this procedure for the sequence, a "local measurement" of the "degree
of correlations" is obtained, i.e. a local measure of the Hurst exponent in the
DFA case. The results indicate that the $H$ exponent value varies
with time. This
is similar in finance to what is observed along $DNA$ sequences\cite{DNA} in
biology where the $H$ exponent drops below 1/2 in so-called non-coding regions.
Doing the same here, we obtain a $local$ Zipf law and $local$ Zipf exponent. To
show a full list of figures or data as a function of $m$ $\in$ [2,8] would
generate  a quite aversive stimulus in the reader, - therefore only
the case $m$
=5 is illustrated here. This value is so chosen within the financial idea
background having motivated this study, i.e. $m=5$  is the (true) length of a
week !

Results for the three FBM signals, with $H$ close to 0.5, like in
financial time
series signals,  are illustrated in Figs. 6-8, considering windows (boxes) of
size 250, 500 and 1000 respectively  moved along the signal. These values
correspond to 1, 2 and 4 year type investment window in finance.   Notice that
the $local$ exponents are usually larger than the corresponding average one, in
some sense corroborating the previous finite size effect analysis results. In
turns it seems that the method is also of interest in order to observe short
range correlation in fractional Brownian motions, and non ergodic properties of
finite size series.

\section{Zipf Analysis of Fractional Brownian Motion + Linear Trend Data}

Next consider a FBM on which a linear trend with amplitude $A_L$ is
$additionally$ superposed (and is equal to zero at the origin), i.e. we add
\begin{equation} y_L(i) =A_L i \end{equation} to  Eq. (4). We have taken values
of $A_L$ ranging from $2^{-16}$ to $2^{-7}$ depending on the FBM
considered.  The
case $k$=2 is only considered at this stage, again as a first step
and having in
mind scales used in other works in econophysics and electronics. The Zipf plots
are shown in Fig.9. For small $H$  a sharp drop is still seen at large $R$
values. The amplitude (and trend) effect is hardly noticed at small $H$, even
with a large trend, while the linearity (on a log-log plot) is interestingly
observed for larger $H$.

The effect of the trend when $m$ is larger than 2,  thus up to $m$=8, as in the
preceding section is illustrated in Fig. 10.  As should be expected
the exponent
$\zeta '$ is rather stable  (= $\zeta_0 '$ ) for a small trend (or slope) $A_L$
and small $m$, but markedly increases when $m$ increases. The $\zeta
'$ stability
is observed apparently below some sort of crossover amplitude
$A_{L,cr}(m)$, ...
depending on $m$. The variation above such a  crossover amplitude
follows a power
law  ($\zeta ' \sim  A_{L}^{\theta_L}$ ) which has been determined.
The values of
the exponent and of the power law variation on both sides of the $A_{L,cr}$ are
given in Table 2, with their error bar $\Delta \theta_L$.

We should point out that there seems to be some difference whether
the FBM signal
is persistent or not. For antipersistent signals, $\zeta '$ is quasi $A_{L}$
independent. The Brownian motion is clearly an $in$ $between$ case.

\section{Conclusions}

We have considered   fractional Brownian motions on which  a linear trend is
superimposed. We have translated the signals each into a text based of two
letters $u$ and $d$, (or 0 and 1 bits)  according to the fluctuations in the
corresponding random walk. We have studied the variation of the Zipf
exponent(s)
giving the relationship between the frequency of occurrence of words
of length $m
\le 8$  made of such ''letters'' for a binary alphabet. We have
searched how the
$\zeta '$ exponent of the Zipf law is influenced by the trend and its
amplitude.
We can distinguish finite size effects, and results depending whether the
starting FBM is persistent or not, depending on the Hurst exponent.
It seems that
$\zeta '$ varies as a power law in terms of $m$.  This seems to be due to the
fact that due to the trend a marked bias occurs in the relative fluctuations,
thus in the word occurrences.  It seems proven, even though it might have been
expected so,  that  a persistent signal short range correlations as analyzed
through the Zipf method is more influenced by a trend than an antipersistent
signal.

In the spirit of the local DFA, we have introduced considerations based on  the
effect of finite sizes of texts, and on the notion of a local (or ''time''
dependent) Zipf law, in order to show their influence on the exponent values.

Finally, coming back to the  Czirok et al.
conjecture,\cite{CSV,czirok,troll} it
seems of interest to display the relationship for both studied
classes of cases,
i.e. (i) the raw FBM, and (ii) the FBM + linear trend. The results
are  shown in
Figs. 11-12.  Even within considerations taking into account finite
size effects
and a poor man statistical analysis, it appears that such a law only holds near
$H=0.5$.

Other considerations are in order showing that many other cases can  still be
considered: first one could wonder more about signal stationarity
effects. Next,
either a non linear (thus like a power law or a moving average)  trend or a
periodic background could be superposed on the raw signal.  Also multiplicative
rather than additive trends could be used.   This should be put in
line with the
remarkable studies of Hu et al. \cite{Hu} on DFA but is a work  an order of
magnitude higher  here because there are mk parameters to consider.
Applications
of the above to financial data will be presented elsewhere
\cite{AusloosBronlet}.

\vskip 0.6cm

{\bf Acknowledgments}

\vskip 0.6cm

We don't thank anyone for stimulating discussions and comments, but the referee
for imposing precision and conciseness. 

\newpage

{\large \bf Figure Captions} \vskip 0.5cm

{\bf Figure 1} --  Six fractional Brownian motions studied in the text with
characteristics summarized in Table 1 characterized by an $H$ exponent

\vskip 0.5cm {\bf Figure 2} -- ($m$,2) Zipf plots of the six FBM for
$2 \le m \le
8$

\vskip 0.5cm {\bf Figure 3} -- Evolution of the $\zeta'$ exponent with $m$ for
different $H$ values

\vskip 0.5cm {\bf Figure 4} -- Evolution of the $\zeta'$ exponent
calculated when
successively removing one by one the less frequently occurring words as a
function of $m$ for different $H$ values

\vskip 0.5cm {\bf Figure 5} -- Zipf law $\zeta'$  exponent $evolution$ as a
function of time $t$ starting from the (first) box containing 100 points, then
for a box containing 200, 300, ...  etc.  points up to 16 000

\vskip 0.5cm {\bf Figure 6} -- Time dependence of the $local$ Zipf
exponent in a
box  of size  $T$ displaced along the FBM with $H$=  0.41; three size boxes are
illustrated : 205, 500 and 1000

\vskip 0.5cm {\bf Figure 7} -- Time dependence of the $local$ Zipf
exponent in a
box  of size  $T$ displaced along the FBM with $H$= 0.47; three size boxes are
illustrated : 205, 500 and 1000

\vskip 0.5cm {\bf Figure 8} -- Time dependence of the $local$ Zipf
exponent in a
box  of size  $T$ displaced along the FBM with $H$=  0.60; three size boxes are
illustrated : 205, 500 and 1000

\vskip 0.5cm {\bf Figure 9} -- ($m$,2) Zipf plots of the six
FBM+linear trend for
$2 \le m \le 8$  with slope  $A_L$ of the linear trend given in insert

\vskip 0.5cm {\bf Figure 10} -- Variation of the $\zeta'$ exponent as
a function
of the slope trend when $m$ is larger than 2 and less than 8

\vskip 0.5cm {\bf Figure 11} -- ''Verification'' of the relationship $ \zeta =
|2H-1|$ for the six FBM

\vskip 0.5cm {\bf Figure 12} -- ''Verification'' of the relationship $ \zeta =
|2H-1|$ for the six FBM + linear trend for different trends

\begin{table}[ht] \begin{center} \caption{ Characteristics of the six raw
fractional Brownian motions (FBM) hereby studied (Fig.1) : Hurst exponent $H$,
error on $H$ through box counting method, trend and bias of the FBM}

\begin{tabular}{|c|c|c|c|c|c|c|}
$H$ & $0.17$ & $0.41$ & $0.47$ & $0.60$ & $0.67$  &  $0.82 $  \\
$\Delta H $ & $0.02 $ & $0.01$ & $0.02$ & $0.01$ & $0 .01$ & $0.02$ \\ \hline
$trend (10^{-5})$ & $ 1.09$ & $ 3.03$ & $ -3.2 $ & $ 2.48$ & $4.51$ & $ -6.22 $
\\
$bias$ & $0.0084$ & $-0.0055$ & $-0.00575$ & $-0.004$ & $0.0143$ & $0.0038$ \\
\hline\hline  \end{tabular}

\end{center} \end{table} \newpage

\begin{table}[h,t,b] \begin{center} \begin{tabular}{cccccccc} \hline
&  $m$  & 4
& 5 & 6 & 7 & 8 \\ \hline 1.$H\sim0.17$ &$\zeta_0 '$        & 0.2311 & 0.2336 &
0.2524 & 0.2753 & 0.2982 \\ \\ 2.$H\sim0.41$ &$\zeta_0 '$        & 0.1005 &
0.1001 & 0.1179 & 0.1304 & 0.1445 \\ &$\theta_L$       & 1.3470 &
1.3325 & 1.2884
& 1.1926 & 1.1412 \\ &$\Delta\theta_L$ & 0.0008 & 0.0013 & 0.0006 & 0.0001 &
0.0004 \\ \\ 3.$H\sim0.47$ &$\zeta_0 '$        & 0.0164 & 0.0231 & 0.0307 &
0.0456 & 0.0725 \\ &$\theta_L$       & 1.0816 & 1.1053 & 1.0817 &
1.0773 & 1.0257
\\ &$\Delta\theta_L$ & 0.0014 & 0.0027 & 0.0017 & 0.0016 & 0.0004 \\ \\
4.$H\sim0.60$ &$\zeta_0 '$        & 0.2743 & 0.2773 & 0.2783 & 0.2816
& 0.2854 \\
&$\theta_L$       & 0.7392 & 0.7220 & 0.7162 & 0.7095 & 0.6943 \\
&$\Delta\theta_L$ & 0.0012 & 0.0015 & 0.0015 & 0.0015 & 0.0012 \\ \\
5.$H\sim0.67$ &$\zeta_0 '$        & 0.4658 & 0.4554 & 0.4501 & 0.4541
& 0.4617 \\
&$\theta_L$       & 0.7939 & 0.7653 & 0.7252 & 0.6824 & 0.6381 \\
&$\Delta\theta_L$ & 0.0054 & 0.0039 & 0.0025 & 0.0010 & 0.0010 \\ \\
6.$H\sim0.82$ &$\zeta_0 '$        & 0.5381 & 0.5215 & 0.5098 & 0.5056
& 0.5083 \\
&$\theta_L$       & 0.8886 & 0.8457 & 0.8048 & 0.7636 & 0.7213 \\
&$\Delta\theta_L$ & 0.0108 & 0.0094 & 0.0077 & 0.0060 & 0.0047 \\ \hline
\end{tabular}

\caption{Values of the exponent $\zeta_0 '$ and of its power law variation
$\theta_L$  as a function of the slope of  the trend on both sides of the
$A_{L,cr}$ } \label{tab 2.4} \end{center} \end{table}

\newpage \begin{figure}[ht] \begin{center} \leavevmode \epsfysize=8cm
\epsffile{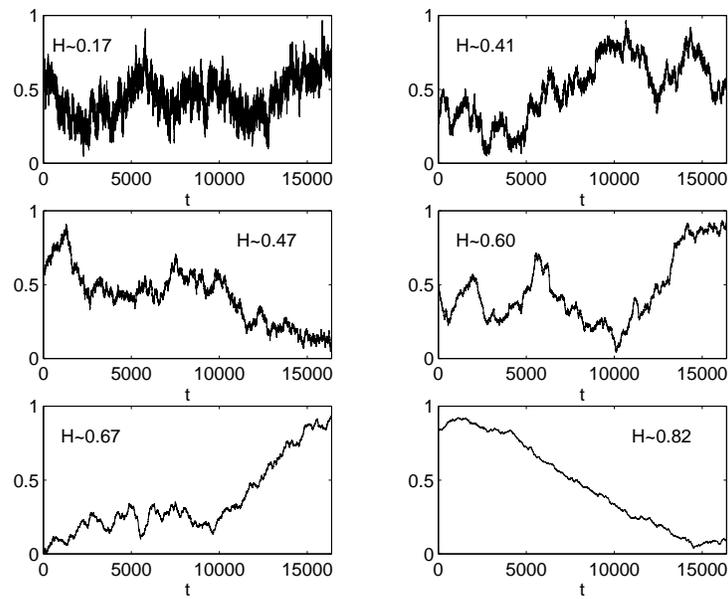} \caption{ Six fractional Brownian motions
studied in the
text with characteristics summarized in Table 1 characterized by an
$H$ exponent}
\end{center} \end{figure}

\begin{figure}[ht] \begin{center} \leavevmode \epsfysize=8cm
\epsffile{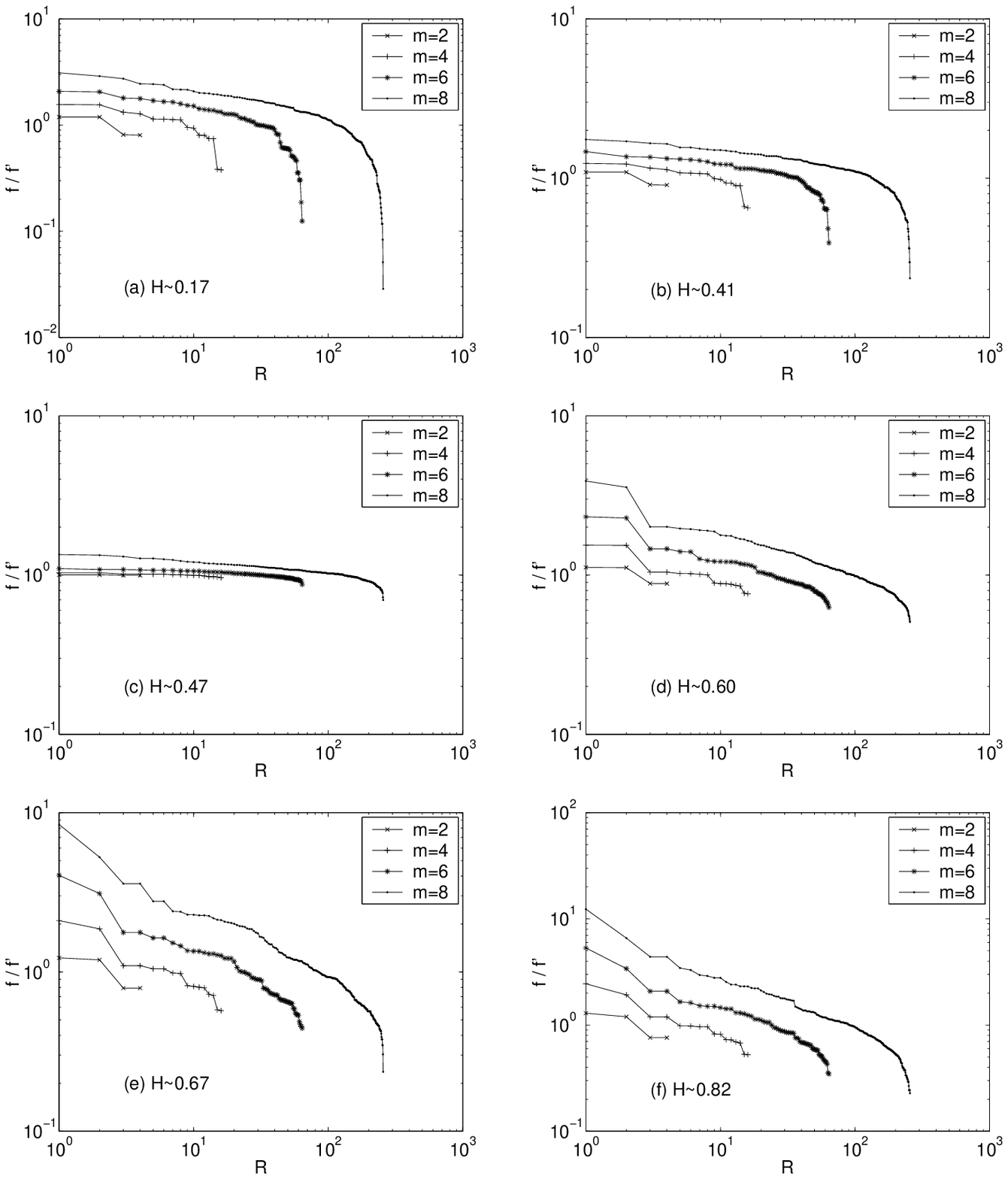} \caption{($m$,2) Zipf plots of the six FBM
for $2 \le m
\le 8$} \end{center} \end{figure}

\begin{figure}[ht] \begin{center} \leavevmode \epsfysize=8cm
\epsffile{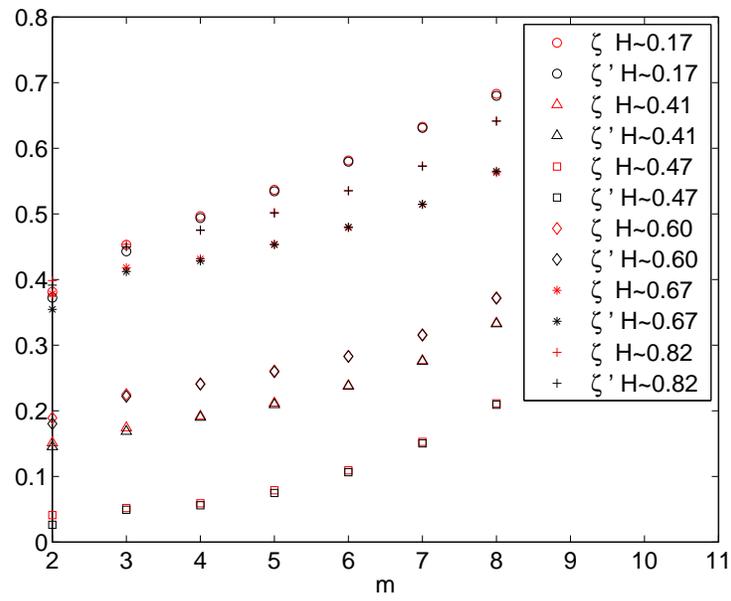} \caption{Evolution of the $\zeta'$ exponent
with $m$ for
different $H$ values} \end{center} \end{figure}

\begin{figure}[ht] \begin{center} \leavevmode \epsfysize=8cm
\epsffile{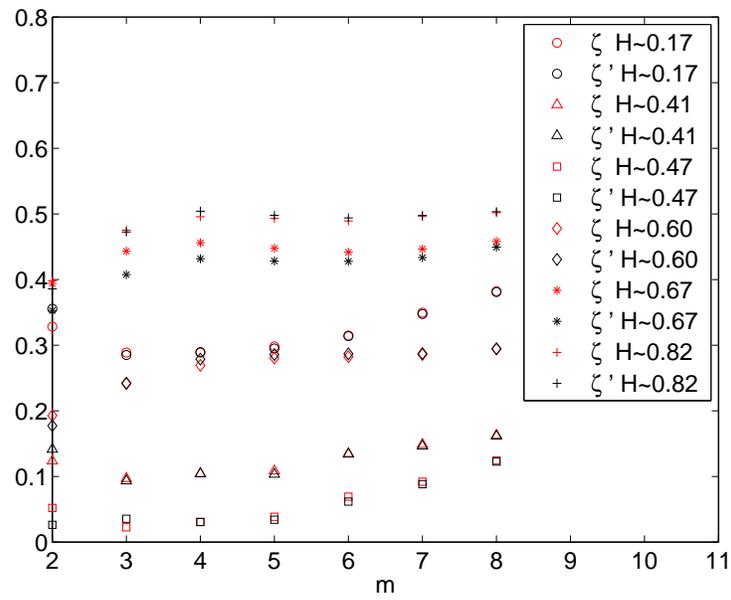} \caption{Evolution of the $\zeta'$ exponent calculated
when successively removing one by one the less frequently occurring words as a
function of $m$ for different $H$ values} \end{center} \end{figure}

\begin{figure}[ht] \begin{center} \leavevmode \epsfysize=8cm
\epsffile{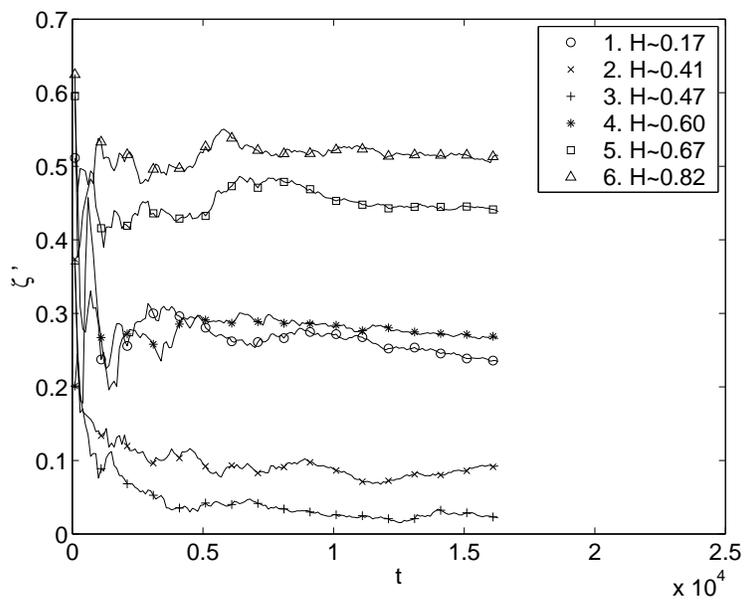} \caption{Zipf law $\zeta'$  exponent $evolution$ as a
function of time $t$ starting from the (first) box containing 100 points, then
for a box containing 200, 300, ...  etc.  points up to 16 000 } \end{center}
\end{figure}

\begin{figure}[ht] \begin{center} \leavevmode \epsfysize=8cm
\epsffile{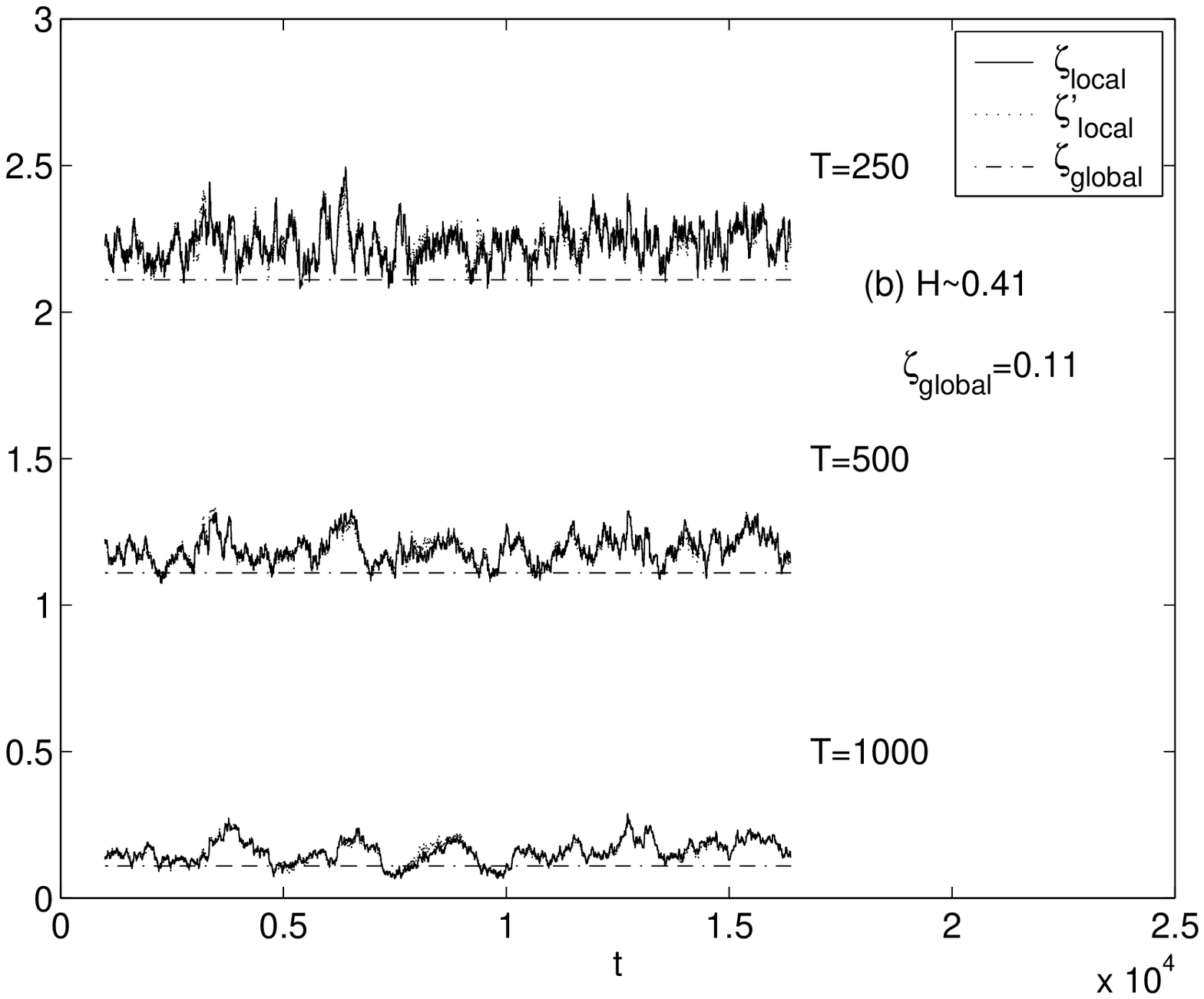} \caption{Time dependence of the $local$ Zipf
exponent in
a box  of size  $T$ displaced along the FBM with $H$=  0.41; three
size boxes are
illustrated : 205, 500 and 1000} \end{center} \end{figure}

\begin{figure}[ht] \begin{center} \leavevmode \epsfysize=8cm
\epsffile{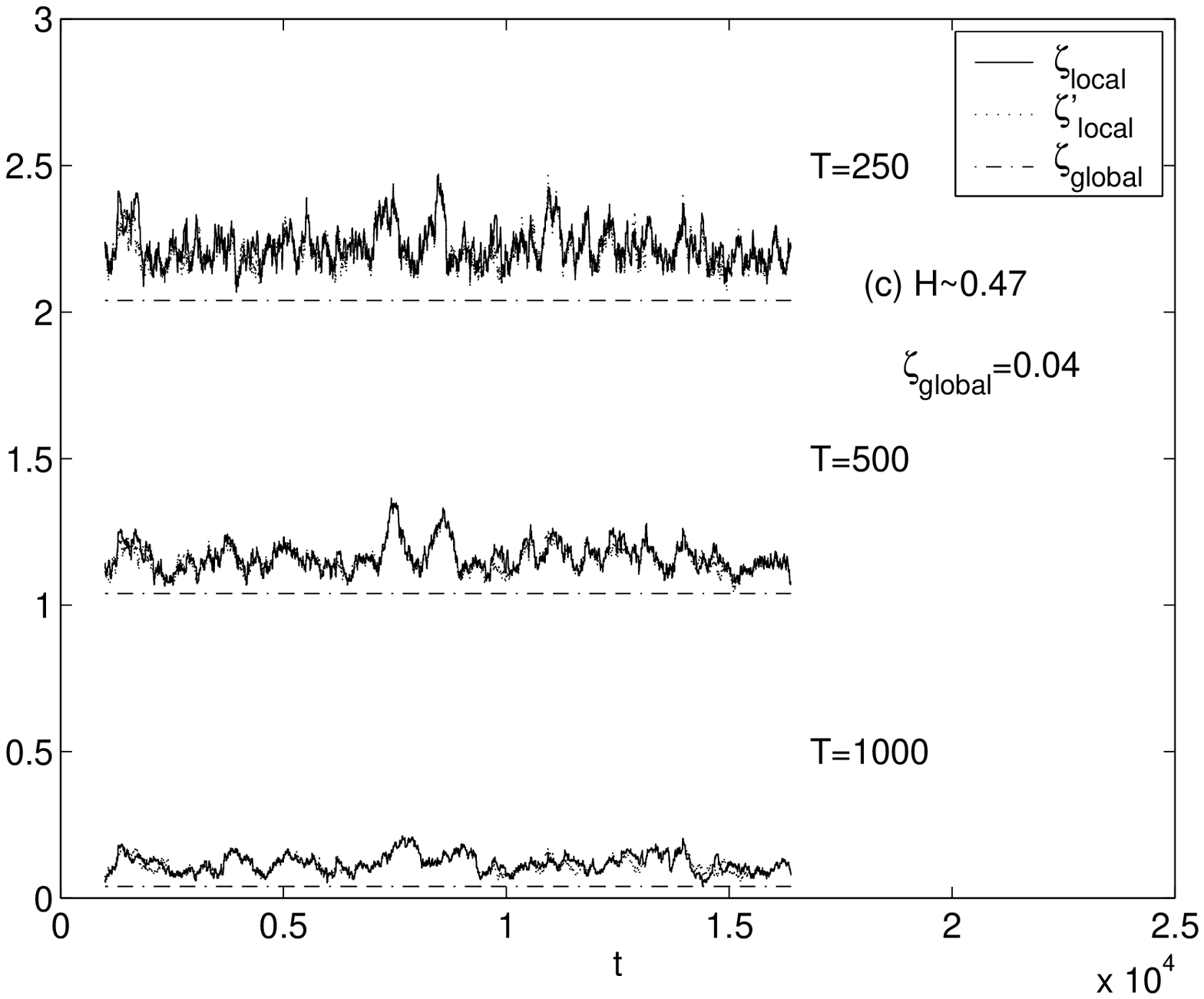} \caption{Time dependence of the $local$ Zipf
exponent in
a box  of size  $T$ displaced along the FBM with $H$= 0.47; three
size boxes are
illustrated : 205, 500 and 1000} \end{center} \end{figure}

\begin{figure}[ht] \begin{center} \leavevmode \epsfysize=8cm
\epsffile{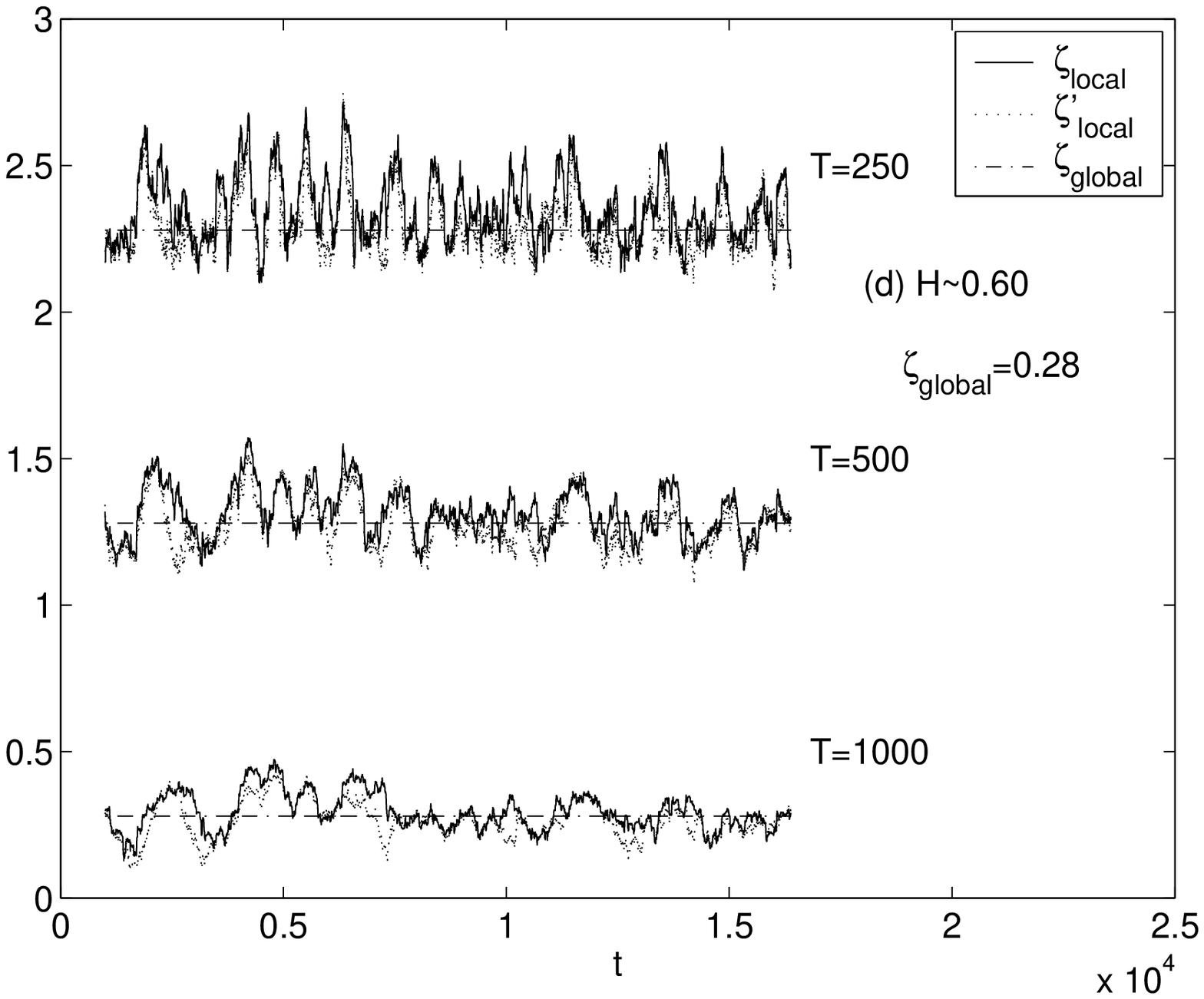} \caption{Time dependence of the $local$ Zipf
exponent in
a box  of size  $T$ displaced along the FBM with $H$=  0.60; three
size boxes are
illustrated : 205, 500 and 1000} \end{center} \end{figure}

\begin{figure}[ht] \begin{center} \leavevmode \epsfysize=8cm
\epsffile{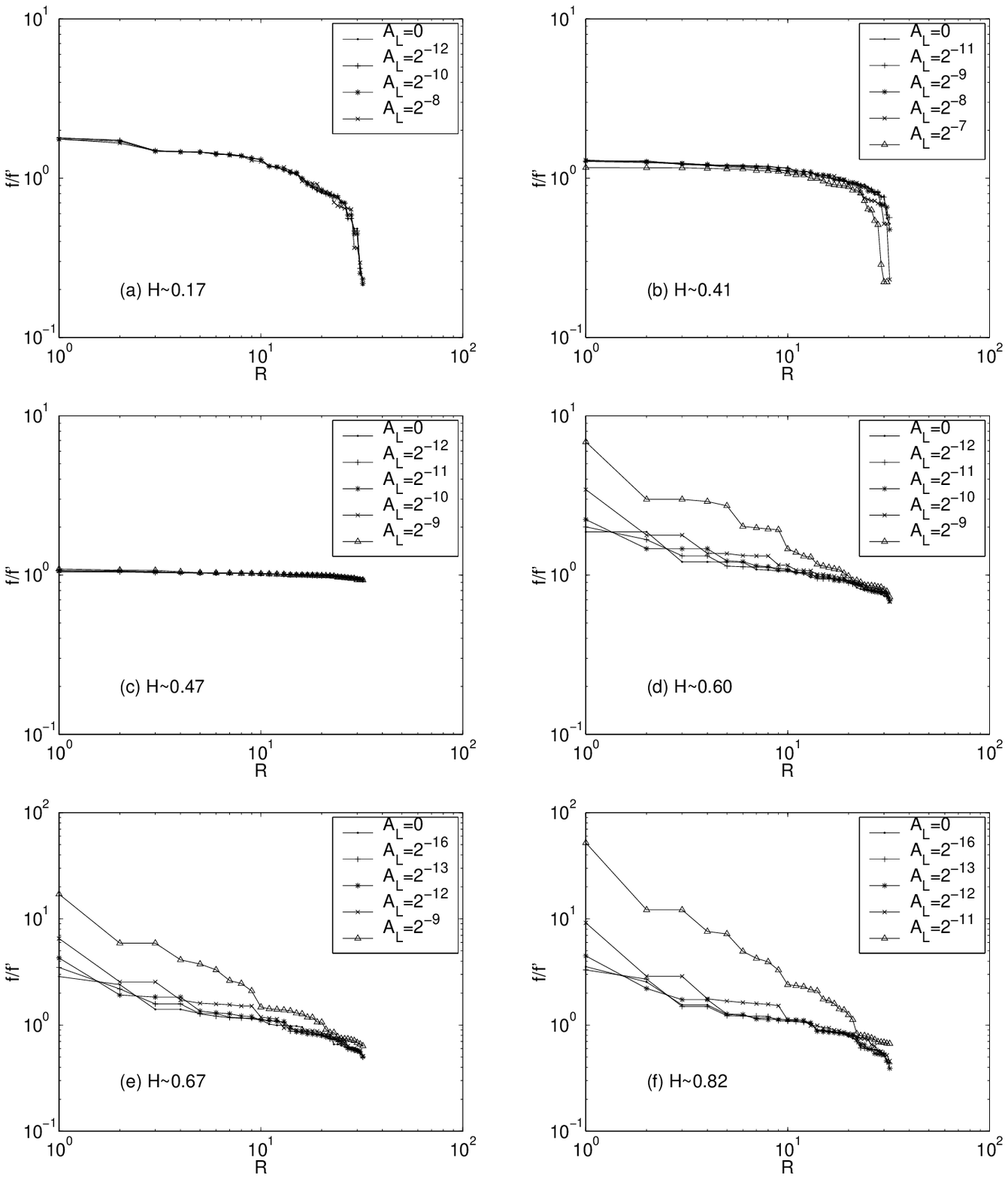} \caption{($m$,2) Zipf plots of the six
FBM+linear trend
for $2 \le m \le 8$  with slope  $A_L$ of the linear trend given in insert}
\end{center} \end{figure}

\begin{figure}[ht] \begin{center} \leavevmode \epsfysize=8cm
\epsffile{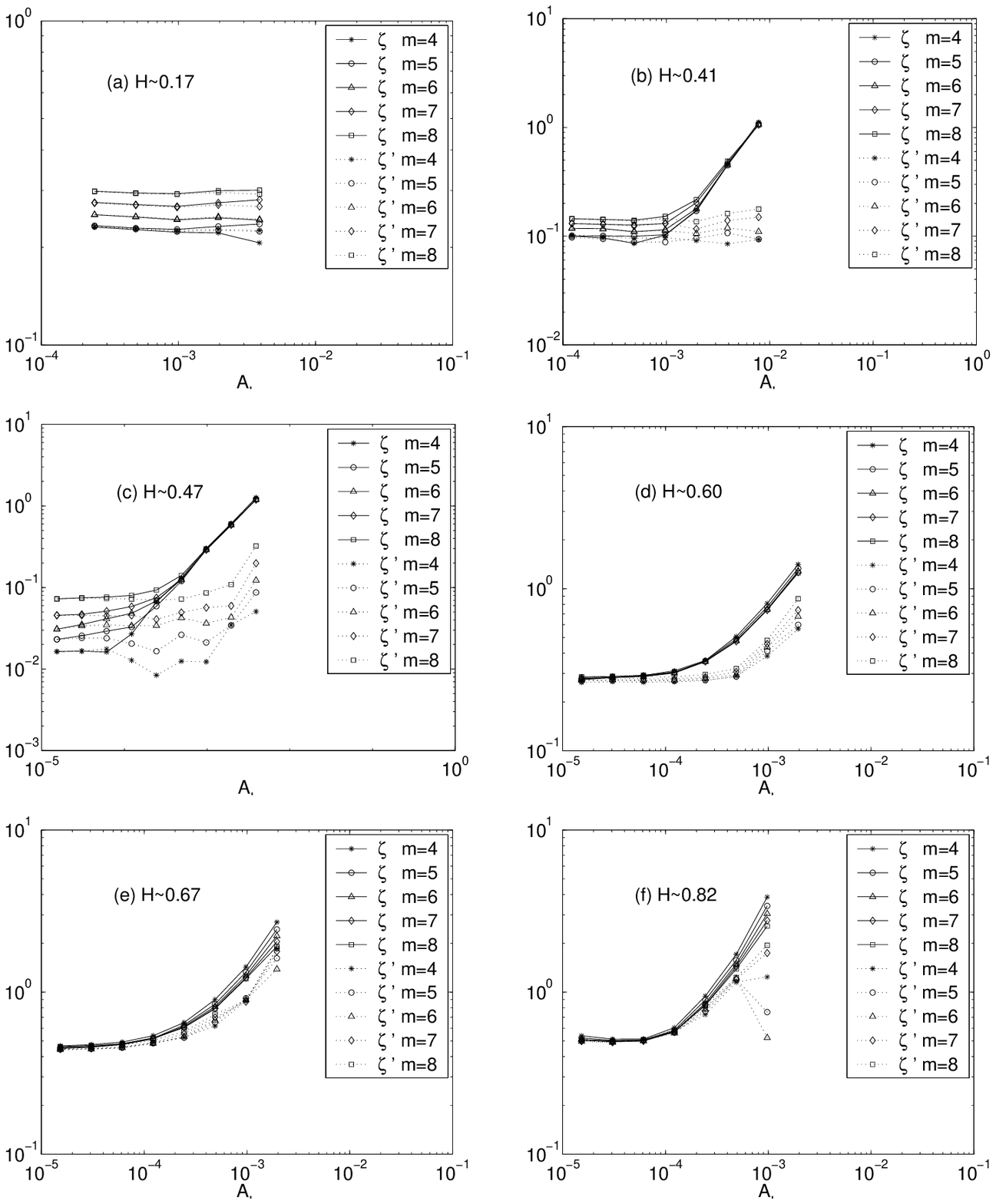} \caption{ Variation of the $\zeta'$ exponent as a
function of the slope trend when $m$ is larger than 2 and less than 8}
\end{center} \end{figure}

\begin{figure}[ht] \begin{center} \leavevmode \epsfysize=8cm
\epsffile{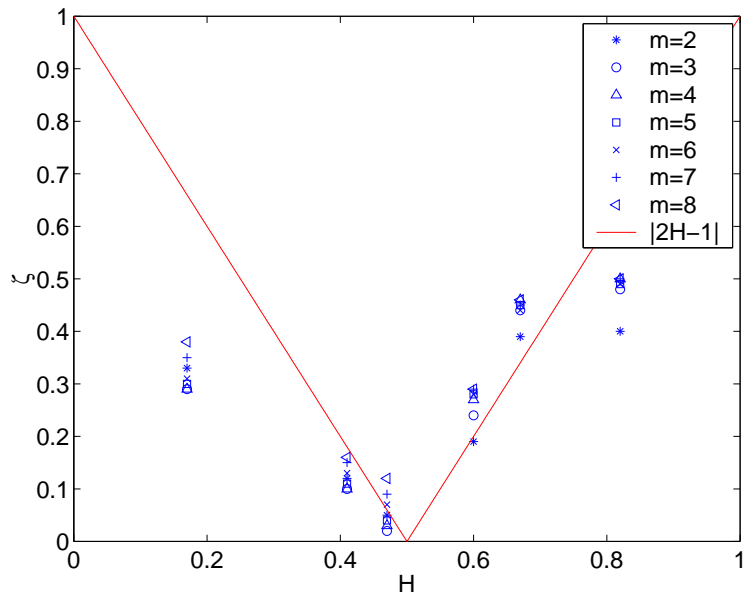} \caption{''Verification'' of the
relationship $ \zeta =
|2H-1|$ for the six FBM} \end{center} \end{figure}

\begin{figure}[ht] \begin{center} \leavevmode \epsfysize=8cm
\epsffile{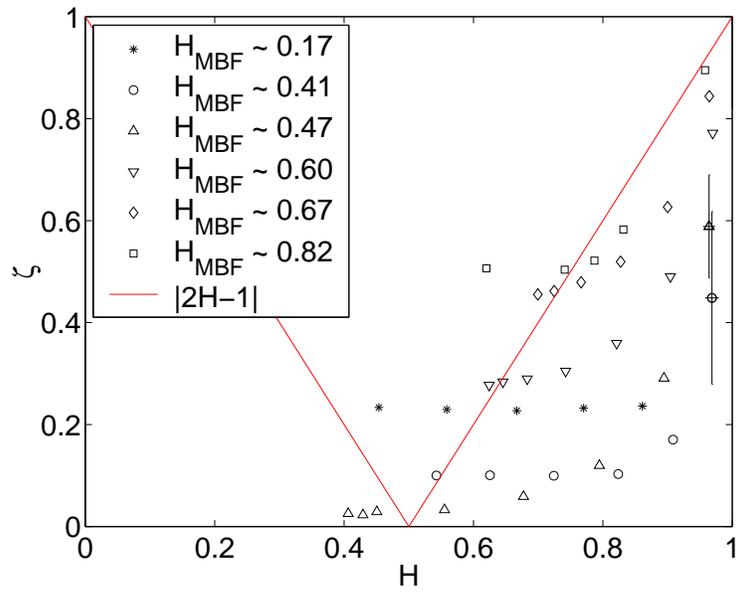} \caption{''Verification'' of the
relationship $ \zeta =
|2H-1|$ for the six FBM + linear trend for different trends} \end{center}
\end{figure}

\end{document}